\documentclass[twocolumn,aps,10pt,floatfix]{revtex4-2}
\usepackage{amsmath, amssymb}
\usepackage{graphicx}
\usepackage{xcolor,ulem}

 \usepackage{siunitx}
\newcommand{\be}{\begin{equation}}
\newcommand{\ee}{\end{equation}}
\newcommand{\bee}{\begin{eqnarray}}
\newcommand{\eee}{\end{eqnarray}}

\begin{document}

\title{Mobility in immersed granular materials upon cyclic loading}

\author{Tanvir Hossain}
\author{Pierre Rognon}
\email[]{pierre.rognon@sydney.edu.au}
\affiliation{Particles and Grains Laboratory, School of Civil Engineering,\linebreak
	The University of Sydney, Sydney, NSW 2006, Australia}

\date{\today}


\begin{abstract}
We study the mobility of objects embedded in an immersed granular packing and subjected to cyclic loadings. In this aim, we conducted experiments using glass beads immersed in water and a horizontal plate subjected to a cyclic uplift force. Tests performed at different cyclic force frequencies and amplitudes evidence the development of three mobility regimes whereby the plate stays virtually immobile, moves up steadily or slowly creeps upwards. Results show that steady plate uplift can occur at lower force magnitudes when the frequency is increased. We propose an interpretation of this frequency-weakening behaviour based on force relaxation experiments and on the analysis of the mobility response of theoretical visco-elasto-plastic mechanical analogue. These results and analysis point out inherent differences in mobility response between steady and cyclic loadings in immersed granular materials. 
\end{abstract}

\maketitle

\section{Introduction}

Objects embedded in granular materials can sustain a non null external force without significantly moving. When this force exceeds a threshold, they start moving continuously through the packing. This basic property arises from the ability of granular materials to deform elastically or to flow like a liquid depending on the level of shear stress they are subjected to \cite{andreotti2013granular}. 

A usual test to measure this force threshold involves moving an initially still object at a constant, slow velocity and monitoring the external force required to drive its motion. In dense packings, this force increases to a maximum $F^{qs}_0$ at small object displacements, while the granular packing deforms mostly elastically. It then decreases as the granular material deforms plastically around the object, a process necessary to enable its progression. $F^{qs}_0$ corresponds to the maximum drag force the granular packing can develop to hinder the object motion. The superscript \textit{qs} stands for \textit{quasi-static}, meaning that this force is measured at an object velocity that is low enough not to influence it  \cite{candelier2009creep,harich2011intruder,kolb2013rigid,seguin2019hysteresis,takada2020drag}. Reciprocally, $F^{qs}_0$ corresponds to the maximum external force the object can statically sustain without continuously moving. Several studies have measured and modelled this maximum drag force in dry granular materials. The consensus is that it follows a frictional law: $F^{qs}_0$ is proportional to the average packing pressure at the object depth, and to its surface area projected in the direction of motion \cite{albert1999slow,albert2000jamming,albert2001granular,hill2005scaling,gravish2010force,costantino2011low,ding2011drag,guillard2013depth,takada2020drag}. Other influential factors include the strength of the packing characterised by an internal friction angle \cite{das2013earth},  the shape of the object \cite{dyson2014pull,askari2016intrusion,giampa2018effect} and the grain size \cite{sakai1998particle,athani2017grain,costantino_starting_2008}.

The presence of interstitial water significantly affects this force threshold. This was evidenced in experiments involving dragging a rod horizontally \cite{allen2019effective} or uplifting a plate in immersed glass beads \cite{hossain2020PRF}. Firstly, buoyancy reduces the effective pressure in the granular packing \cite{ilamparuthi2002experimental,ravichandran2008study}, which in turns reduces $F^{qs}_0$. Secondly,  the maximum drag force $F_0$ acting on the object become rate dependent, and increases linearly with its velocity as: 

\be \label{eq:F0}
F_0 \approx F^{qs}_0 + m v
\ee

\noindent In Ref. \cite{hossain2020PRF}, we proposed that this rate-dependence results from a Darcy-flow mechanism, whereby water flows through the virtually still granular packing. This yielded a scaling law for the viscous coefficient $m$ in terms of the object size $B$, packing permeability $K$, and fluid viscosity $\eta$: $m \propto \eta B^3/K$, with the permeability being controlled by the packing solid fraction $\nu$ and grain size $d$ according to the Kozeni-Carman model \cite{carman1939permeability,rognon2014explaining}: $K \propto  d^2(1-\nu)^3/\nu^2$. However, the drag force component  $m v$ is not static: it gradually relaxes to zero if the plate is stopped, following a dynamics similar to a Maxwell visco-elastic relaxation. The time scale $\lambda$ of this relaxation was associated with the viscous coefficient $m$ and an effective  stiffness parameter $k$ as $\lambda \approx m/k$.

The mobility response of the object subjected to cyclic loading is comparatively less understood. Such loadings are characterised by a force amplitude $F_{max}$ and a frequency $f$. The basic expectation is that the object would not move if $F_{max}<F_0^{qs}$ and would move otherwise. However, simulations of cyclic uplift in dry grains showed that the loading frequency influences the onset of motion \cite{athani2018mobility}. At high frequencies, the plate was found to sustain force amplitudes higher than $F_0^{qs}$ without moving. The proposed interpretation for this \textit{frequency-strengthening} is that the granular packing needs a force greater than $F_0^{qs}$ to plastically deform, but also requires that force to be sustained for a long enough period of time to allow for elementary inertial motion of grains.

The presence of water may induce different internal dynamics. For instance, shaking an immersed granular material is known to lead to the phenomenon of liquefaction, whereby grains become suspended, and the packing strength vanishes \cite{youd2001liquefaction,kayen2013shear,clement2018sinking}. However,  whether and how cyclic loadings affect the object mobility in immersed granular packing remains to be established.

The purpose of this Paper is to address this gap. In this aim, we conducted experimental results showing the mobility response of a horizontal plate embedded into immersed grains, and subjected to a cyclic upward force. Tests involve prescribing different loading magnitude and frequency, and monitoring the resulting vertical motion of the plate.  These results are then used as a basis for analysing the plate dynamics, which is made by comparing it to the cyclic loading response of a visco-elasto-plastic mechanical analogue. The Paper is structured as follows. The experimental method is presented in Section \ref{sec:method}, the measured mobility response in Section \ref{sec:mobility} and the analysis of its underlying dynamics in Section \ref{sec:analysis}. 

\begin{figure}[!tb]
{\includegraphics[width=1\columnwidth]{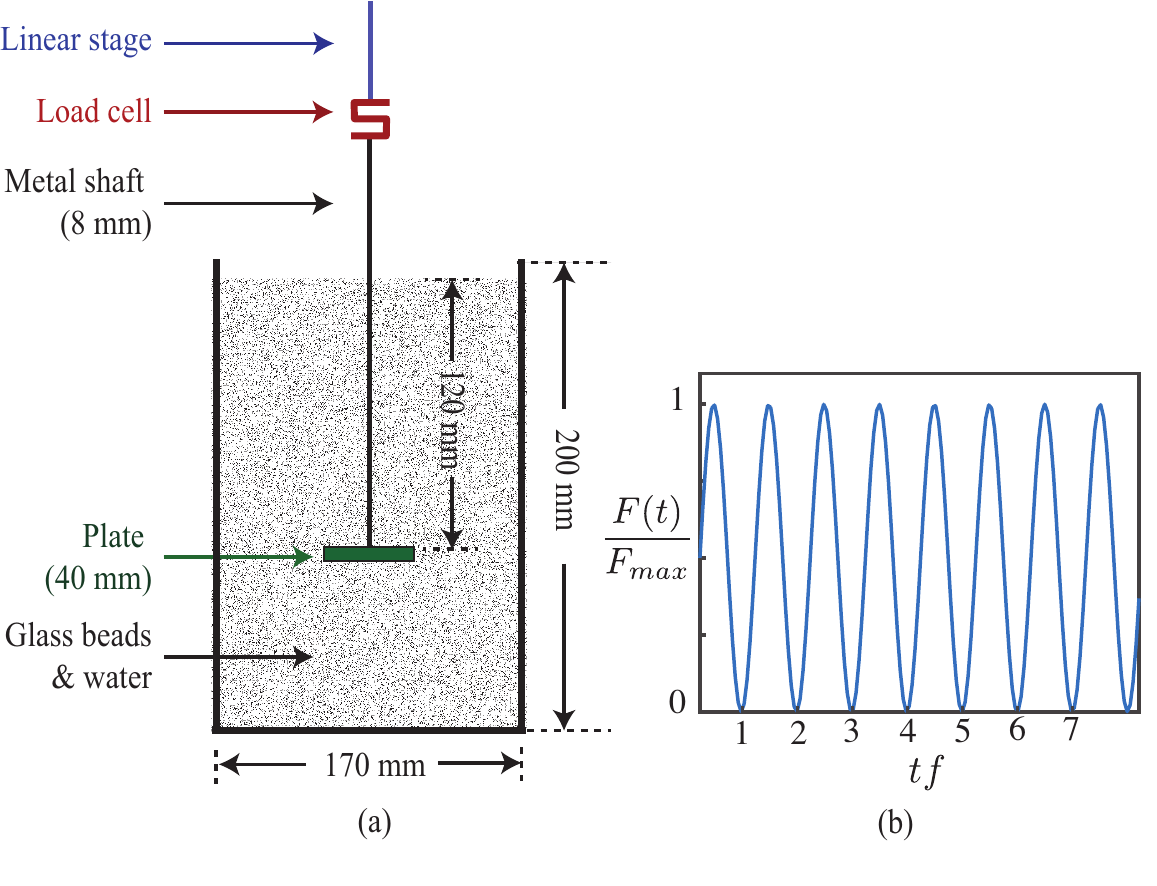}}
\caption{Cyclic uplift tests. (a) Illustration of the experimental set-up showing the initial location of the plate. (b) Time evolution of the cyclic uplift force $F(t)$ applied to the plate via the shaft, which is characterised by an amplitude $F_{max}$ and a frequency $f$ as per Eq. (\ref{eq:force}).}
\label {exp_setup}       
\end{figure}

\begin{figure*}[tb!]
{\includegraphics[width=1\textwidth]{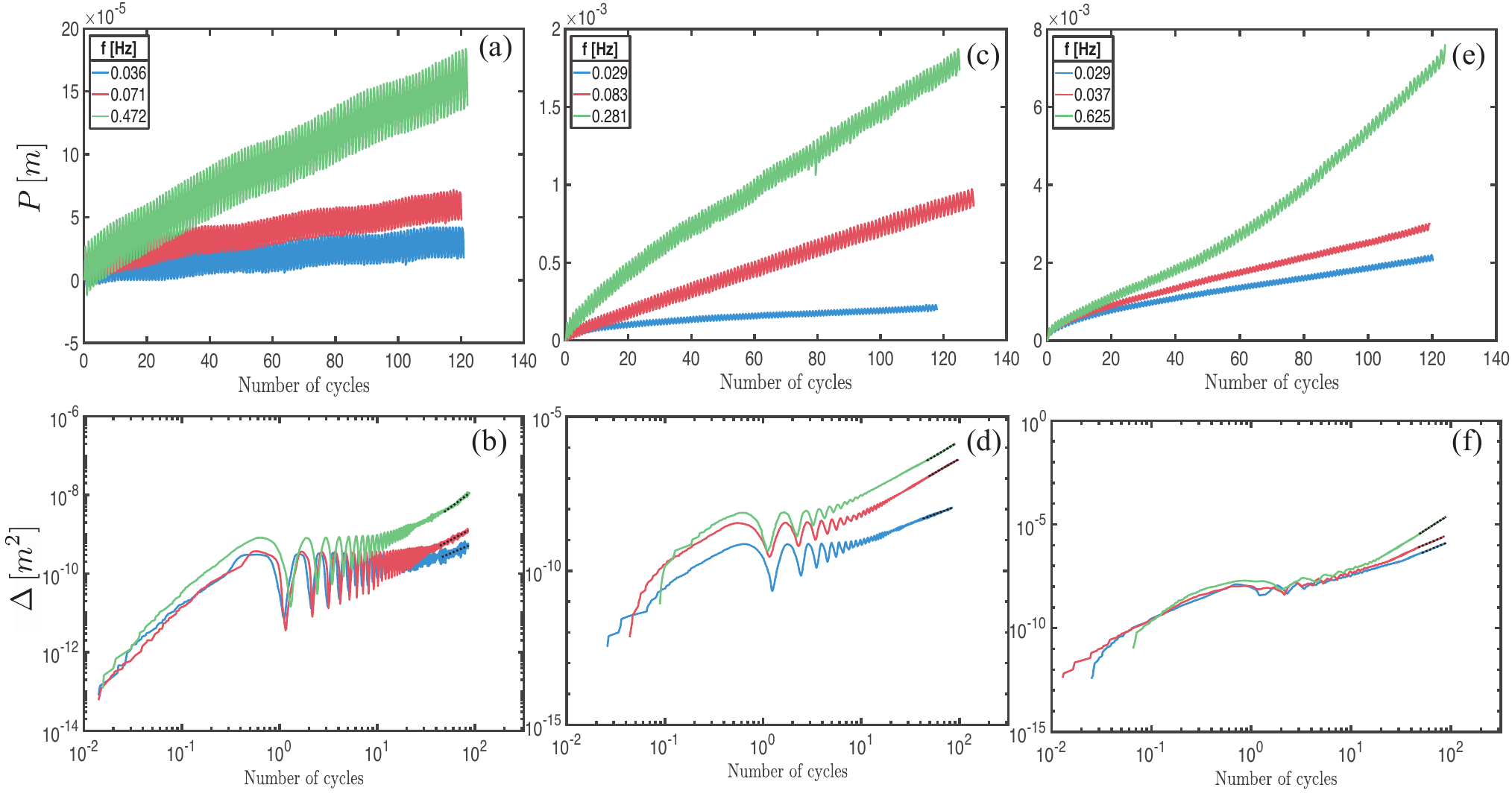}}
\caption{Evolution of the plate displacement $P$ and mean square displacement $\Delta$ measured at different loading magnitudes and frequencies.  $F_{max}$ =  \SI{4}{N} (a, b), \SI{6}{N} (c, d) and \SI{8}{N} (e, f). The frequencies $f$ are given in the legends in (a), (c), and (e). On (b,d,f), the black lines show the power law fits of the mean square displacement according to (\ref{eq:alpha}) considering a number of cycle $\tau f$ greater than $50$ cycles (see text). Values of the resulting power $\alpha$ are shown on figure \ref{fig:alpha}. }
\label {fig:MSD}       
\end{figure*}

\section{Experimental cyclic uplift tests} \label{sec:method}

\subsection{Materials and set-up}

We conducted the cyclic uplift tests using the experimental setup presented in Figure \ref{exp_setup}a. This setup was previously used in Refs \cite{hossain2020rate,hossain2020PRF} to measure drag forces in dry or immersed granular materials. It comprises a container filled with glass beads, in which a circular disk-shaped plate is embedded at designated depth $H$. 
The plate has a circular cross-section of diameter $B$ = \SI{40}{mm} and is \SI{4}{mm} thick. The container diameter is \SI{170}{mm}, which is about four times larger than the plate diameter. The mean grain diameter of the glass beads used in the experiments is d= \SI{300}{\micro\meter}, with a polydispersity of $\pm 10\%$. Grains are fully immersed in water. All tests are performed with an initial plate depth of $H$ = \SI{120}{mm} corresponding to an embedment ratio $\frac{H}{B} = 3$. The plate is placed at an equal distance from the vertical edge of the container. 

The plate is driven by a loading frame comprised of a stainless steel shaft connecting the plate to a load cell, which is itself fixed to a ball-screw linear stage powered by a DC servo-motor. The system enables us to either control the plate displacement and monitor the force or vice-versa.  Data logging is made at a frequency of \SI{100}{Hz}.

In Refs. \cite{hossain2020PRF}, we presented a detailed study of this setup which led to the following conclusions. Firstly, the elastic stiffness of the shaft and load cell leads to negligible deformations within the force range we consider (typically less than \SI{50}{N}). Secondly, the maximum drag force $F_0$ of the shaft alone, measured by performing an uplift test without a plate, is of the order of \SI{1}{N}, with is an order of magnitude lower than the resistance measured with the plate. Thirdly, the mode of preparation which involves pouring the mixture of grains and water below the plate, placing the pate and then pouring the mixture of grain and water above the plate is repeatable: repeating uplift tests led to consistent values of maximum drag force $F_0$ with variations of the order of $\pm 10\%$. Lastly, finite-size effects of the container do not affect these results, unless the plate is deliberately placed off-centred at a distance shorter than \SI{20}{mm} from the container edge.

\subsection{Uplift cyclic loading tests}

The cyclic uplift tests involve periodically applying a vertical force on the plate via the shaft. The loading frame controls this force $F(t)$ at any point in time and monitors the plate vertical displacement $P(t)$. We chose a sinusoidal shape for the force cycles (see Figure \ref{exp_setup}b):

\be \label{eq:force}
F(t) = \frac{F_{max}}{2} \left( 1+ \sin \left( 2\pi f t -\frac{\pi}{2}\right) \right).
\ee

\noindent  Accordingly, the applied force varies from $0$ at the beginning of the test $t=0$ to a maximum of $F_{max}$. The force is always directed upward, and the plate is never pushed back downward during a cycle. The cyclic force is characterised by two parameters: its magnitude $F_{max}$, which is the maximum force applied to the plate during a cycle, and the cycle frequency $f$. Accordingly, the rate at which the force varies is
$\dot F(t) = 2\pi f F_{max} \cos \left( 2\pi f t -\frac{\pi}{2}\right)$,  which reaches a maximum of: 

\be
\dot F_{max} = 2 \pi f F_{max}
\ee

\section{Measuring mobility regimes}\label{sec:mobility}

We conducted a series of cyclic tests at different force magnitudes and frequencies. $F_{max}$ was varied from \SI{4}{N} to \SI{10}{N} and $f$ from \SI{0.02}{s^{-1}} to \SI{0.86}{s^{-1}}. Tests were stopped after $120$ cycles.The reasons for exploring these ranges will become apparent in the next section. This section focuses on empirically evidencing different mobility regimes.

The basic quantity characterising the mobility of the plate a during the cyclic uplift test is its vertical displacement $P$ defined as:

\be
P(t) = y(t)-y(t=0)
\ee 

\noindent where $y(t)$ is the vertical position of the plate at time $t$. Figure \ref{fig:MSD} shows some examples of such measured trajectories, by plotting the measured plate displacement as a function of the number of cycles $n$; the relation between $n$ and $t$ is $n=ft$. It appears that the plate trajectory strongly depends on the loading magnitude and frequency. Higher frequencies and higher magnitudes seem to lead to larger displacements for a given number of cycles. Measured displacements after 120 cycles range over two orders of magnitude: from \SI{30}{\micro\meter} ---which is only a tenth of a grain diameter--- to \SI{7}{mm}. Importantly, the evolution of the displacement appears to be qualitatively different between tests, as it exhibits different degrees of non-linearity. Consequently, a quantity such as $P(n)/n$ where $n$ the number of cycles would not be relevant to characterise the degree of mobility as it would be dependent on the chosen number of cycle $n$.

\begin{figure}[tb!]
{\includegraphics[width=1\columnwidth]{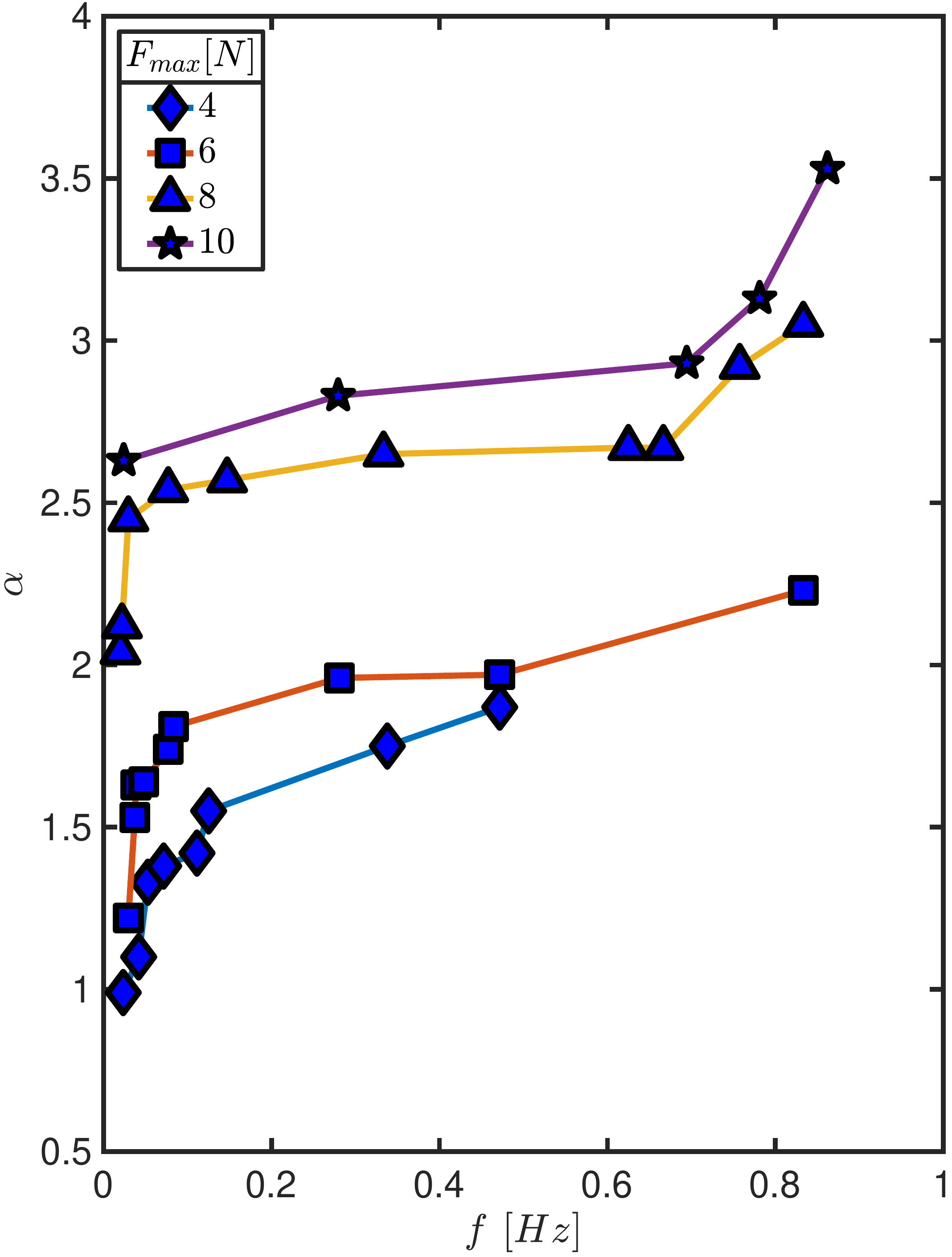}}
\caption{Degree of mobility $\alpha$ measured for all tests including different loading frequencies $f$ and magnitudes $F_{max}$. $\alpha$ is calculated by fitting the plate mean square displacement $\Delta$ by (\ref{eq:alpha}) considering $\tau f$ greater than $50$ cycles.}
\label {fig:alpha}       
\end{figure}

\begin{figure}[htb!]
 {\includegraphics[width=1\columnwidth]{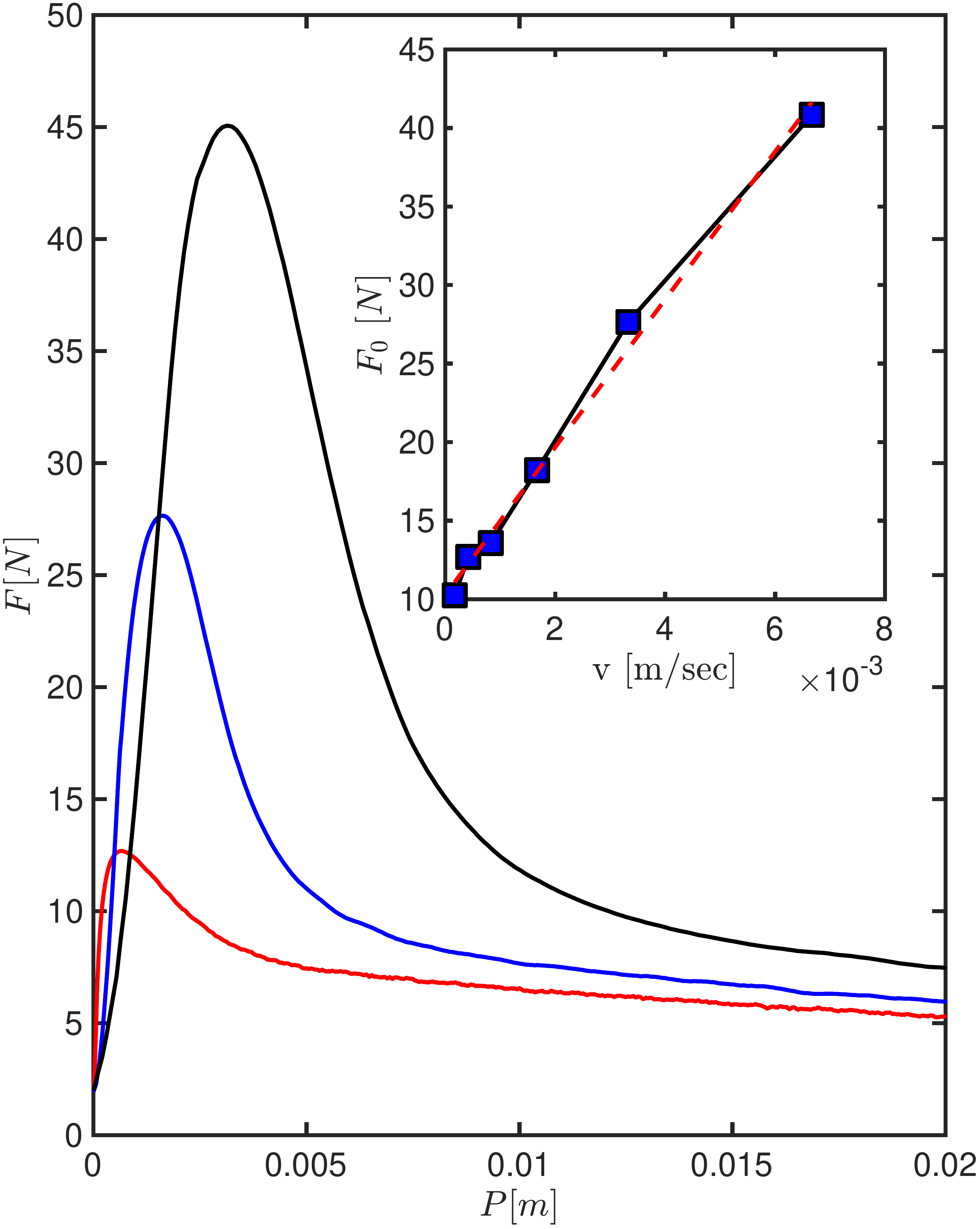}}
\caption{Drag forces in steady uplift tests performed at constant velocities. (Main)  Evolution of drag force $F$ with the plate displacement $P$ for three tests performed at different velocities. (inset) Maximum drag force $F_0 = \max \left(F(P)\right)$ measured in tests performed at different velocities; the line represent the best fit of $F_0(v)$ by the linear function (\ref{eq:F0}), which is obtained for $F_0^{qs}$ = \SI{8.3}{N} and $m$ = \SI{4500}{N/(m/s)}.}
\label{fig4}      
\end{figure}

\begin{figure*}[tb!]
 {\includegraphics[width=1.4\columnwidth]{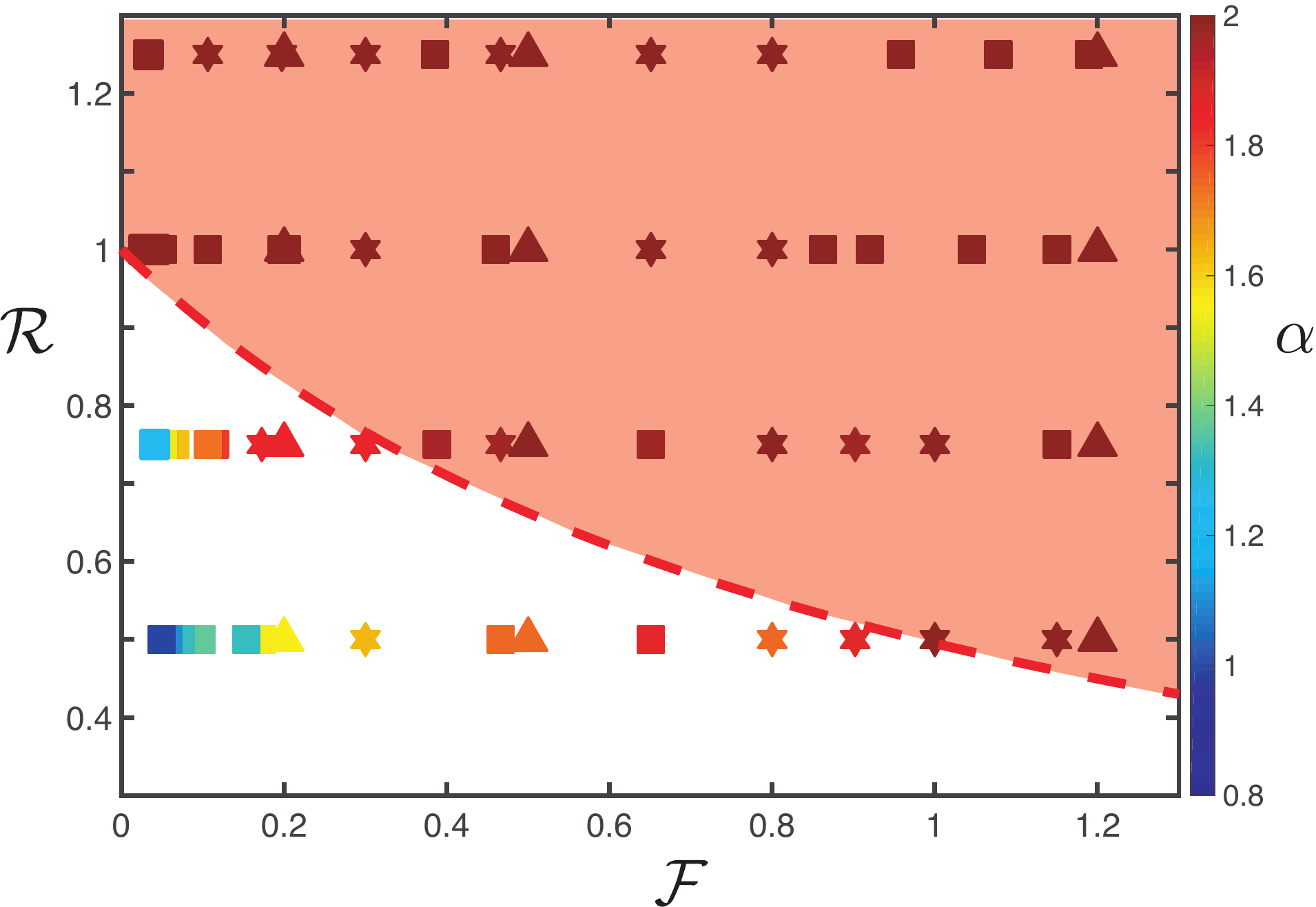}}
\caption{Mobility regime map. Symbols correspond to individual tests performed with a value of loading frequency and magnitude corresponding to a value of the dimensionless numbers $\mathcal{R}$ and $\mathcal{F}$, respectively. Symbols are coloured according to the value of the degree of mobility quantified by the power $\alpha$. Squares ($\blacksquare$) correspond to experimental data.  Stars ($\bigstar$) and triangles ($\blacktriangle$) correspond to results obtained using the visco-elasto-plastic analogue (see figure \ref{fig8}); stars are obtained using the same parameter as the experimental data while triangles are obtained using a quasi-static maximum drag $F^{qs}_0$twice as big and a relaxation time $\lambda$ twice as low (see text). The red dashed line is the theoretical limit $\mathcal{R}^c(\mathcal{F})$ for failure regime (highlighted in red) according to the proposed model in Eq. (\ref{eq:Rc}).}
\label{fig5}      
\end{figure*}

\begin{figure*}[htb!]
 {\includegraphics[width=1\textwidth]{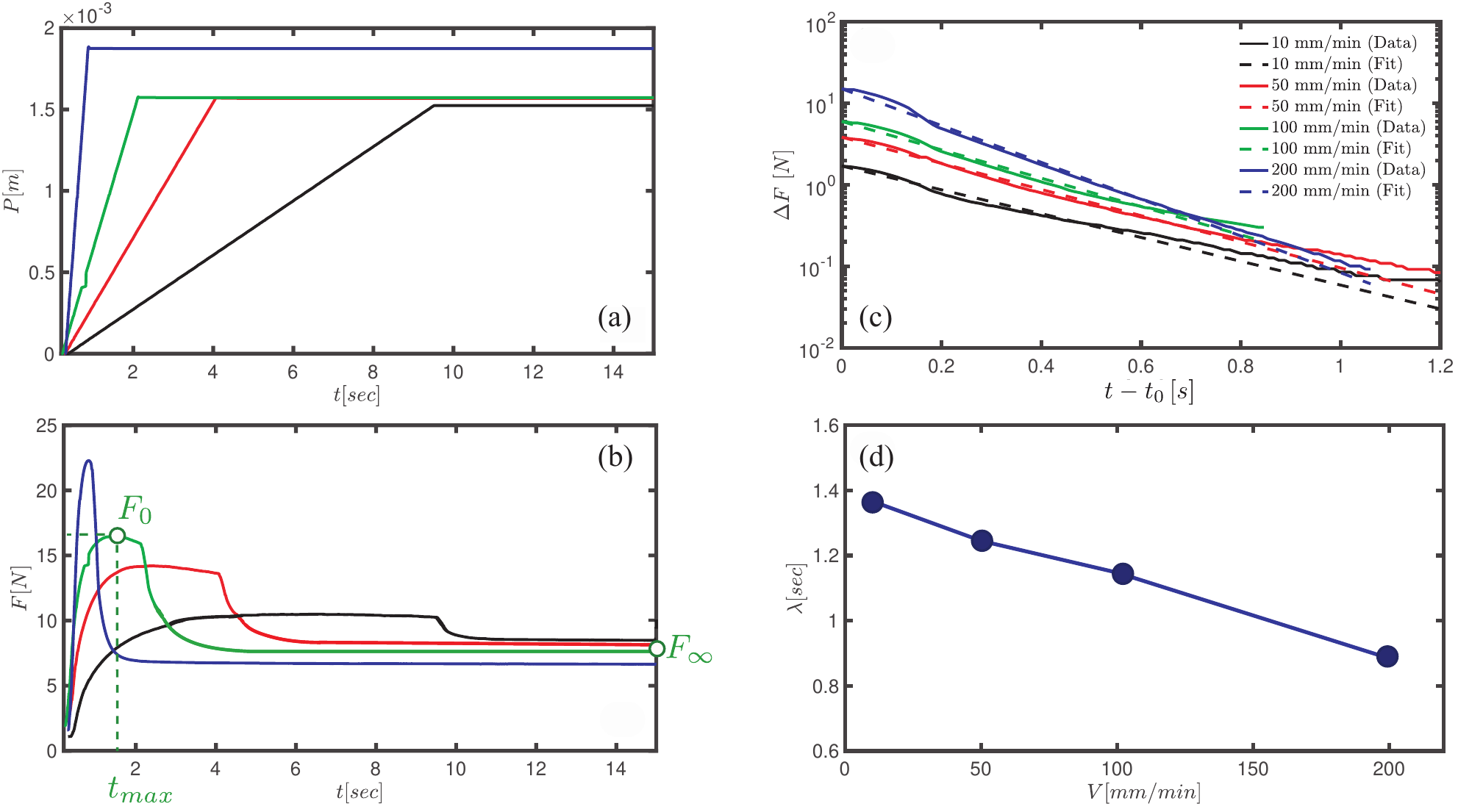}}
\caption{ Drag force relaxation experiments. (a-c) Example of four tests performed at different initial velocities $v$ (see legend in (c)). (a) Controlled plate displacement $P(t)$: the plate is moved at a constant velocity $v$ and stopped shortly after the maximum drag force is reached, at $t= t_0$. (b) Corresponding drag force $F(t)$, showing the relaxation at $t>t_0$. (c) Force relaxation $\Delta F(t-t_0) = F(t) - F(t=t_0)$ and exponential decay fits using (\ref{eq:relax}) with the time scale $\lambda$ as a free parameter. (d) Fitted value of the relaxation time scale $\lambda$ as a function of the plate's initial velocity. }
\label{fig6}      
\end{figure*}

\begin{figure*}[htb!]
 {\includegraphics[width=1\textwidth]{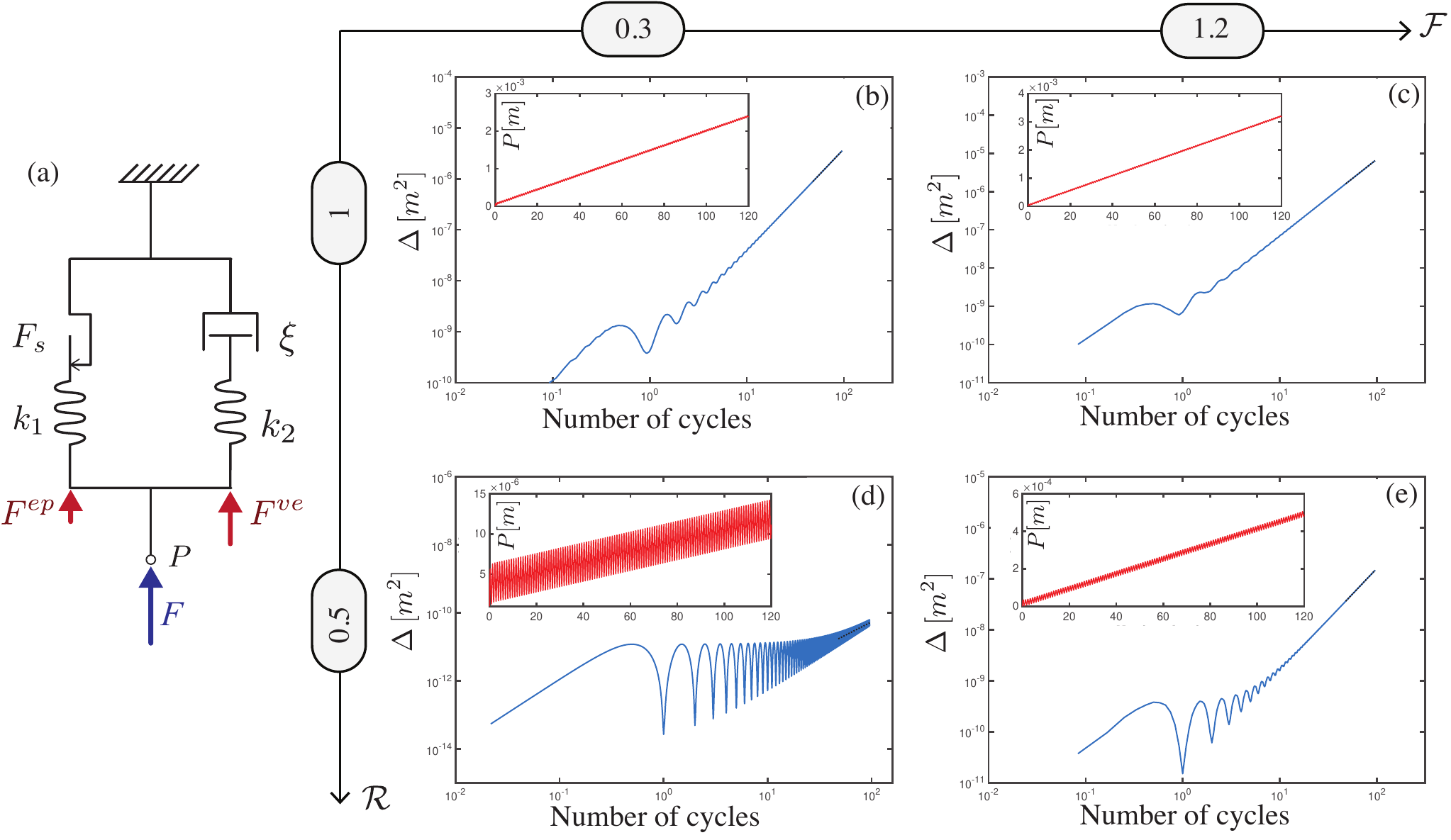}}
\caption{Visco-elasto-plastic analogue. (a) Network diagram showing a visco-elastic and a viscoplastic element in parallel. The force $F$ applied at point $P$ is shared between these two elements: $F=F^{ep}+F^{ve}$. (b-e) Displacements and mean square displacements simulated by applying a cyclic loading (\ref{eq:force}) to the analogue in (a) and using a frequency dependent slider threshold $F_s$ (see text); the four panels correspond to different values of loading amplitude $\mathcal{R}$ and frequency  and $\mathcal{F}$ as shown on the figure. (b,c,e) lead to a failure regime. The associated degree of mobility $\alpha$ are shown in figure \ref{fig5}.}
\label{fig8}      
\end{figure*}

\begin{figure}[h!]
 {\includegraphics[width=1\columnwidth]{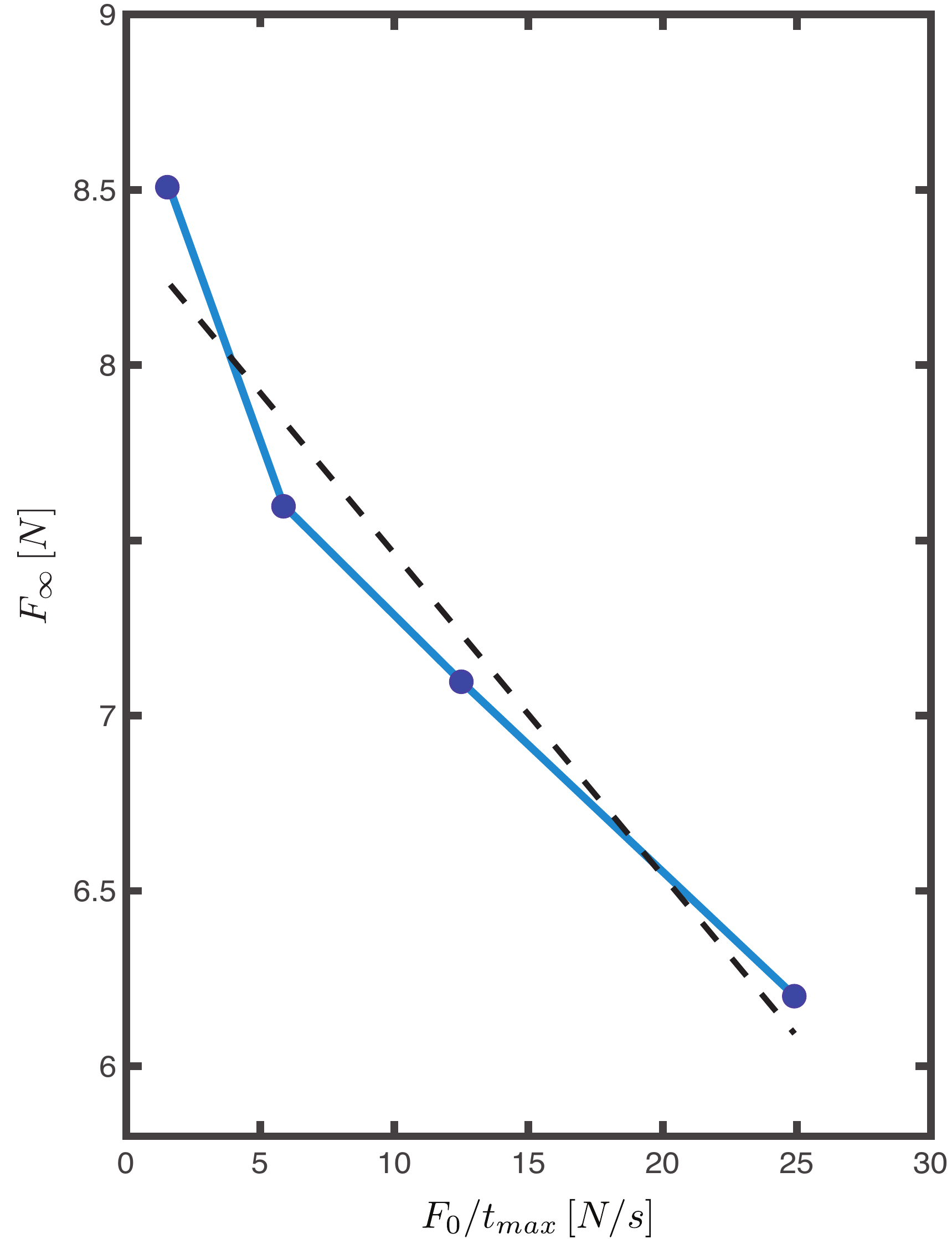}}
\caption{Rate weakening of the drag force measured in relaxation tests: ultimate drag force $F_{\infty}$ as a function of the initial rate of force increase $F_{0}/t_{max}$ (see figure \ref{fig6}b). Symbols correspond to experimental data and the dashed line is the linear fit (\ref{eq:Finf_v}).}
\label{fig7}      
\end{figure}

To quantify the degree of mobility for each tests, we therefore followed the approach introduced in Ref. \cite{athani2018mobility}. It is based on a metric derived form the average mean square displacement $\Delta$ of the plate, which is defined as:

\begin{equation}
\Delta (\tau ) = \frac{1}{N}\sum_{n=1}^{N}\left | y (\tau + t_n) -y(t_n) \right |^2.
\label{eq:msd}
\end{equation}

\noindent $y(t_n)$ denotes the position of the plate at a reference time $t_n$, $y (\tau + t_n)$ represents the position of the plate after a time increment $\tau$ and $N$ denote the total number of reference time $t_n$, which were taken at random during the first half of the tests.

Figure \ref{fig:MSD} shows the evolution of the mean square displacement measured from tests conducted with different magnitudes and frequencies.  Like with dry granular materials \cite{athani2018mobility}, $\Delta$ exhibits two phases: it first oscillates at short values of $\tau$ corresponding to a few cycles. At larger values of $\tau$, it increases at various rates. In a first approximation, this increase is consistent with a power-law:

\be \label{eq:alpha}
\Delta (\tau ) \propto \tau^{\alpha}
\ee 

\noindent In the context of random walk and diffusion, the interpretation of such power law is the following  \cite{griffani2013rotational,ariel2015swarming,combe2015experimental,kharel2017vortices}. A power $\alpha=1$ corresponds to normal diffusive behaviour and a random walk trajectory. A power $\alpha >1$ corresponds to a super-diffusive behaviour: $\alpha =2$ corresponds to a plate moving at a constant velocity, while $\alpha >2$ means that the plate is accelerated. $\alpha <1$ corresponds to a sub-diffusive behaviour. In Ref. \cite{athani2018mobility}, three mobility regimes were defined according to the value of $\alpha$:

\begin{enumerate}
\item A regime of confinement for $\alpha < 1$. In this regime, the plate does not significantly move after each cycle.
\item A regime of creep for $ 1 \leqslant \alpha \leqslant 2$. In this regime, the plate incrementally moves up by small (random) steps after each cycle.
\item A failure regime for $\alpha > 2$.  In this regime, the plate consistently moves up through the packing after each cycle.
\end{enumerate}

Figure \ref{fig:alpha} reports the values of the power $\alpha$ calculated for all tests by fitting the measured mean square displacement $\Delta (\tau )$ with  Eq. (\ref{eq:alpha}). To exclude the initial oscillating part of the mean square displacement, the fits consider values of $\tau f$ greater than $50$ cycles. It appears that the power $\alpha$ significantly varies with both the loading magnitude and frequency. Expectedly,  larger force magnitudes lead to a higher degree of mobility. Unexpectedly, results evidence that higher frequencies yield to a higher degree of mobility. We refer this effect as \textit{frequency-weakening}. It contrasts with the opposite effect of frequency strengthening reported with dry grains \cite{athani2018mobility}.

\section{Physical mechanisms}\label{sec:analysis}

This section discusses some physical mechanisms that can possibly control the observed mobility response to cyclic loading, including frequency-weakening. The discussion first focuses on two elementary experimental tests aiming at measuring the quasi-static maximum drag force $F_0^{qs}$ and the dynamic drag force relaxation. It is then guided by the analysis of the response of a visco-elasto-plastic analogue subjected to such loadings.

\subsection{Quasi-static maximum drag force $F_0^{qs}$}

We first seek to compare the magnitude of the cyclic loading to the drag forces developing in steady loading. In this aim, we repeated uplift tests similar to those presented in Ref. \cite{hossain2020PRF}, which involves  driving up the plate at a constant velocity $v$. Figure \ref{fig4} shows that the drag force $F(P)$ increases to a maximum $F_0$, and then decreases. Measurement of $F_0(v)$ evidences a linear increase such as (\ref{eq:F0}).
 With the plate size, grain size and plate depth considered here, the best linear fit of the measured $F_0$ is obtained for $F_0^{qs}$ = \SI{8.3} and $m$ = \SI{4500}{N/(m/s)}.

These results imply that applying an external force lower than $F_0^{qs}$ should not be sufficient to sustain the plate motion. Conversely, applying a force larger than $F_0^{qs}$ should lead to the plate moving continuously upward. Extrapolating this expectation to the case of cyclic loading predicts that the plate should reach a failure regime if and only if the applied force exceeds $F_0^{qs}$ at some point in a cycle, which translates into $F_{max}>F_0^{qs}$. To test this, Figure \ref{fig5} shows a map of the measured mobility regime as a function of the dimensionless number:

\be
\mathcal{R} = \frac{F_{max}}{F_0^{qs}}
\ee

\noindent and a dimensionless number $\mathcal {F}$ proportional to the loading frequency, which will be defined below. \noindent Figure \ref{fig4} confirms that a failure response is consistently observed when $\mathcal{R}>1$. Failure response is indeed observed for all tests performed with ($\mathcal{R}\geqslant 1$). However, failure responses also occur at lower magnitudes ($\mathcal{R}<1$) when the frequency is increased. This suggests that the criteria delineating the failure response would not only involve $\mathcal{R}$ but also the frequency $f$.
  
\subsection{Drag relaxation time $\lambda$}

We now seek to identify a characteristic time scale for the drag force, which could be used as a comparison point for the loading frequency. The presence of water is known to lead to a visco-elastic dynamics  when granular materials are subjected to a step increase in stress \cite{cassar2005submarine,rognon2010internal,rognon2011flowing,ikeda2019universal}. To translate this behaviour in the context of object mobility, we performed drag force relaxation tests whereby the plate is uplifted at a constant velocity $v$ until the maximum force $F_0(v)$ is reached, and is subsequently stopped. Figure \ref{fig6} shows that the drag force $F(t)$ typically relaxes exponentially in time toward a non-null value $F_\infty = F(t \to \infty)$ after the plate is stoped, according to:

\be \label{eq:relax}
\frac{F(t>t_0)-F(t=t_0)} {F_\infty-F(t=t_0) }=  1 - e^{-\frac{t}{\lambda}} 
\ee

\noindent $t_0$ is the time at which the plate is stopped, and $\lambda$ is a characteristic time scale of the force decay. Figure \ref{fig6} reports the values of $\lambda$ obtained by fitting measured force relaxation $F(t>t_0)$ with (\ref{eq:relax}). Fits are performed with $\lambda$ as a sole free parameter while the values of $F(t=t_0)$ and $F_\infty$ are determined from the experimental data; $F_\infty$ is approximated by the last value of force reading available at the end of a test. It appears that the relaxation time is of the order of $\lambda \approx \SI{1.1}{s}$, and has a weak dependency on the initial velocity $v$.
This relaxation time enables us to form a dimensionless number for characterising the loading frequency:

\be\label {eq:F}
\mathcal {F} = f \lambda
\ee

\noindent $\mathcal {F}$ compares the drag relaxation time to the period of a force cycle. Accordingly, $\mathcal {F} \lesssim 1$ corresponds to a regime where the drag force has enough time to relax within a cycle. Conversely, $\mathcal {F} \gtrsim 1$ corresponds to a regime where the drag force does not have enough time to relax between cycles. 

\subsection{Visco-elasto-plastic analogue}

In Ref. \cite{hossain2020PRF}, we found that both the linear increase in drag force with velocity (\ref{eq:F0}) and its relaxation dynamics (\ref{eq:relax}) could be captured by the mechanical analogue presented on figure \ref{fig8}a.
This visco-elasto-plastic analogue defines a dynamic for the position of the point $P$ as a function of the applied force $F$. It considers that this force is split between two elements that experience the same deformation: an elasto-plastic element comprised of a spring (spring constant $k_1$) and a slider (force threshold $F_s$), and a Maxwell visco-elastic element comprised of a spring (spring constant $k_2$) and a viscous dashpot (viscous constant $\xi$).

To capture both (\ref{eq:F0}) and (\ref{eq:relax}), the slider threshold needs to equates $F_0^{qs}$, the viscous damper needs to equate the coefficient $m$, $\xi=m$ and the spring constant $k_2$ must be $k_2 = \xi \lambda$. The elasto-plastic element is thought to represent the force transmitted from the plate to the granular matrix and the elasto-plastic deformations they induce. The visco-elastic element is thought to account for the force transmitted via the fluid, which governs some pore pressure gradient and induces a Darcy flow. $k_2$ is may be seen as an effective stiffness corresponding to the elastic deformation of the packing induced by the Darcy flow. 

However, driven by a cyclic force such as (\ref{eq:force}), this analogue predicts a regime of failure only if the slider slides. This means that the applied force $F(t)$ must be at least greater than $F_s$ at some point in the cycle. This would translate into a frequency-independent failure criterion $\mathcal{R}>1$. Accordingly, this analogue cannot readily capture the observed frequency-weakening.

\subsection{Frequency weakening} 

We now seek to enrich the visco-elasto-plastic analogue to capture the observed frequency-weakening. The reasoning stems from the experimental relaxation tests, which show that the drag force relaxes to a value that  can be lower than than the quasi-static maximum drag: $F_\infty<F_0^{qs}$. Figure \ref{fig7} shows that the value of $F_\infty$ decreases with the average rate of increase of the drag force $<\dot F> =  \frac{F_0}{t_{max}}$, where $t_{max}$ is the period of time during which the force rises from $0$ to $F_0$. In a first approximation, we propose to model this dependency by a linear function:

\be \label{eq:Finf_v}
F_\infty \approx F_0^{qs} - a  <\dot F>
\ee

\noindent where the coefficient $a$ has a dimension of time. With our experimental conditions, we found that $a\approx 0.09s$. This may be seen as a process of rate-induced weakening of the drag force: the faster it increases, the lower it eventually relaxes too. 

Drawing this behaviour into the context of cyclic loading, we assume that the effective slider threshold is reduced due to the increase in applied force, and that it is given by $F_\infty$ as defined in (\ref{eq:Finf_v}). We further assume that the value of $ <\dot F>$ is proportional to the maximum rate of force increase during a cycle:  $<\dot F> = \beta 2\pi f F_{max}$ where $\beta$ is a numerical constant. This leads to a frequency dependent slider threshold given by:

\be \label{eq:Fsf}
F_s(f) = F_0^{qs} - a \beta  2\pi f F_{max}
\ee

\noindent Considering that a failure regime can develop provided that the force during a cycle exceeds $F_s(f)$ defines an critical force amplitude
$F^c_{max} = F_0^{qs} - a \beta  2\pi f F^c_{max}$. This failure criteria can be expressed in terms of a critical dimensionless number $\mathcal{R}^c = F^c_{max}/F_0^{qs}$ as: 

\be \label{eq:Rc}
\mathcal{R}^c (\mathcal{F}) = \frac{1}{1+ \frac{a}{\lambda} \beta  2\pi \mathcal{F}}
\ee

\noindent Accordingly, cyclic loadings which magnitude amplitude and frequency lead to  $\mathcal{R}> \mathcal{R}^c (\mathcal{F})$ can develop a failure response.
Figure \ref{fig4} shows that this criterion qualitatively captures the frequency-weakening measured experimentally using a constant $\beta = 1.6$. 

To assess the ability of the visco-elasto-plastic analogue to capture the frequency-weakening effect on the mobility regime, we simulated its response to cyclic loadings at different  frequencies and amplitudes.
The simulation involves integrating the displacement $P(t)$ using a first order forward difference scheme of the following set of equations:

 \bee
\dot{P} &=& \frac{ \dot F^{ve}}{k_2} +\frac{ F^{ve}}{\xi} \label{eq:Fve}\\
F^{ve} &=& F(t) - F^{ep}\\
   \dot F^{ep}&=& 
\begin{cases}
   k_1 \dot P \text { if } F^{ep}<F_s\\
    0    \text{ otherwise}
\end{cases} \label{eq:Fep}\\
P(t=0) &=& 0;\\
 F^{ep}(t=0)&=& F^{ve}(t=0) =0
\eee
 
 \noindent The simulations use as an input an external force $F(t)$ given by (\ref{eq:force}) and a frequency-dependent slider threshold given by (\ref{eq:Fsf}). 
 
 A first set of simulation was conducted using parameters measured experimentally: $F_0^{qs}$ = \SI{8.3}{N} and $\xi = m$ = \SI{4500}{N/(m/s)}. The stiffness $k_2$ is defined as $k_2 = \xi/\lambda$ to match the experimental relaxation time $\lambda$, which is considered to be constant and equal to \SI{1.1}{s}. The stiffness $k_1$ is the only parameter that is not defined by experimental data. It is chosen to be proportional $k_1 = 0.5 k_2$ in these simulations. We checked that its value did not significantly affect the analogue dynamics.
 
Figure \ref{fig8} shows that the simulated trajectories and mean square displacement are quantitatively similar to the experimental results. MSD features an oscillating part at short time scale, and a power-law increases at long time scales. We measured the corresponding degree of mobility $\alpha$ by fitting the numerical results by  (\ref{eq:alpha}). Figure \ref{fig5} shows that a failure regime then develops according to the frequency-dependent criteria (\ref{eq:Rc}).

In order to assess the scaling with the dimensionless number $\mathcal{R}$ and $\mathcal{F}$, we performed a second set of simulations using different quasi-static maximum drag and viscous parameter: $F_0^{qs}$ = \SI{16.6}{N} and $\xi = m$ = \SI{4500}{N/(m/s)}, with the same spring constants. Higher values of $F_0^{qs}$ would correspond to a larger or deeper plate, while a larger value of $m$ would correspond to a larger plate or a plate in smaller grains \cite{hossain2020PRF}. This yields a relaxation time $\lambda$ \SI{2.2}{s}. Figure \ref{fig5} shows the degree of mobility obtained with these parameters. They evidence that the failure regime is captured in terms of the dimensionless number $\mathcal{R}$ and $\mathcal{F}$ according to (\ref{eq:Rc}) even with different values of $F_0^{qs}$ and $\lambda$.

\section{conclusion}

This study points out some important features characterising the mobility in immersed granular material in water upon cyclic loading.

Taking the example of a plate subjected to a cyclic uplift force, we found that the plate mobility response is strongly dependent not only on the loading magnitude but also on its frequency.
We observed three mobility regimes  called confined, creep and failure, whereby the plate does not move significantly, slowly creeps up or consistently moves up after each cycle. Similar mobility regimes have previously been observed in dry granular materials \cite{athani2018mobility}. However, we find here that the presence of water strongly affects their conditions of occurrence. 

With water, we highlighted a process of frequency-weakening that contrasts with the frequency-strengthening observed with dry grains:  the failure regime can occur at low loading magnitudes when the frequency is high. Remarkably, the plate can thus be pulled out even if the force never exceeds the quasi-static maximum drag $F_0^{qs}$, which defines the onset of failure under steady loadings. We propose to understand this by considering that the strength of the packing is weakened during cyclic loadings, and that the extent of this weakening increases as the frequency is increased. This mechanism is supported by experimental relaxation tests showing decay in residual drag force as its initial rate of growth is increased. This observation is captured by the phenomenological law (\ref{eq:Finf_v}). Translating this effect into the context of cyclic loading led us to establish a frequency-dependent criterion for the onset of a failure regime in Eq. (\ref{eq:Rc}), which captures the experimental observations. 

We further found that a visco-elasto-plastic analogue could qualitatively reproduce these mobility responses, provided that its sliding criterion is dependent on both the loading magnitude and frequency according to  (\ref{eq:Fsf}). This broadens the range of validity of this analogue, which was previously shown to capture both the measured maximum drag force $F_0$ and its relaxation dynamics, as per Eqs. (\ref{eq:F0}) and (\ref{eq:relax}).

These results provide a basis to rationalise the mobility response upon cyclic loading considering objects of differing size and shape, buried at different depth in granular materials of different size. According to our findings, a frequency dependent failure criteria similar to (\ref{eq:Finf_v}) should be expected, taking into account the specific quasi-static maximum drag force $F^{qs}_0$ and its relaxation time $\lambda$. For instance, larger or deeper objects would have a larger $F^{qs}_0$, and finer granular materials would lead to longer relaxation time. Similarly,  one could expect a frequency weakening to occur when the loading is applied laterally or downward.

\bibliographystyle{apsrev4-2}
\bibliography{biblio,biblio2}  

\end{document}